\documentclass[oldversion]{aa}
\usepackage{graphicx}
\usepackage{txfonts}
\usepackage{natbib}

\renewcommand{\d}{\mathrm{d}}

\newcommand{\be}{\begin{equation}}
\newcommand{\ee}{\end{equation}}
\renewcommand{\d}{\mathrm{d}}

\begin{document}

\title{
  An optimal basis system for cosmology:\\ 
  data analysis and new parameterisation
}

\author{
  Matteo Maturi\inst{1}
  \and Claudia Mignone\inst{1}\inst{2}
}

\titlerunning{
  An optimal basis system for cosmology
}

\authorrunning{
  M. Maturi \& C. Mignone
}

\institute{
  Zentrum f\"ur Astronomie der Universit\"at Heidelberg,
  Institut f\"ur Theoretische Astrophysik, Albert-\"Uberle-Str.~2,
  69120 Heidelberg, Germany \and Max-Planck-Institut f\"ur Astronomie,
  K\"onigstuhl~17, 69117 Heidelberg, Germany
}

\date{
  \emph{Astronomy \& Astrophysics, submitted}
} 

\abstract{ We define an optimal basis system into which cosmological
  observables can be decomposed. The basis system can be optimised for
  a specific cosmological model or for an ensemble of models, even if
  based on drastically different physical assumptions. The projection
  coefficients derived from this basis system, the so-called features,
  provide a common parameterisation for studying and comparing
  different cosmological models independently of their physical
  construction. They can be used to directly compare different
  cosmologies and study their degeneracies in terms of a simple metric
  separation.  This is a very convenient approach, since only very few
  realisations have to be computed, in contrast to
  Markov-Chain Monte Carlo methods. Finally, the proposed basis system
  can be applied to reconstruct the Hubble expansion rate from
  supernova luminosity distance data with the advantage of being
  sensitive to possible unexpected features in the data set. We test
  the method both on mock catalogues and on the SuperNova Legacy
  Survey data set.}


\maketitle

\section{Introduction}

In the past decade, various cosmological quantities have been object
of intense observational efforts to build our picture of the universe:
the luminosity distance-redshift relation of type-Ia supernovae
\citep[e.g.][]{RI98.1,PE99.6,AS06.1,KO08.2}, the baryonic acoustic
oscillation \citep[e.g.][]{ES05.1}, the cosmic microwave background
(CMB) power spectrum \citep[e.g.][]{BE02.1,KO08.1}, and the cosmic
shear correlation function \citep[e.g.][]{BE07.1,FU08.1}, etc. All of
these data sets are usually interpreted and explained through a direct
comparison with a specific model, or a class of models as for example
Friedmann cosmologies, which are inevitably based on simplifications
and assumptions. A remarkable example is the equation of state
parameter of dark energy, $w$, whose behaviour is still poorly
understood. Thus, if the adopted model ignores unexpected features
which may actually exist, the results may be largely
misleading. Several authors highlighted the pitfalls that the weak
dependence of the equation of state parameter on the actual
observables produces on the possible conclusions drawn on the
dark-energy properties \citep[e.g.][]{MAO01.1, MAO02.1, BA04.3}.

A model-independent approach, instead, may not be affected by these
limitations. The importance of a model-independent reconstruction of
the cosmic expansion rate from luminosity distance data has been
widely discussed in the past decade. The possibility of reconstructing
the dark-energy potential from the expansion rate, $H(a)$, or from the
growth rate of linear density perturbations, $\delta(a)$, was first
pointed out by \cite{STA98.1}, where the relations between the
observational data and the expansion rate are presented.  

Several different techniques have been developed since then to
appropriately treat the data in order to perform such a reconstruction
\citep[see, e.g.][]{HUT99.1, HUT00.1, TEG02.1, DA03.2, WAN05.1}, all of them
employing a smoothing procedure in redshift bins. A recent
reconstruction technique, which recovers the expansion function from
distance data, has been developed in \cite{SHA06.1} and
\cite{SHA07.1}, making use of data smoothed over redshift with
Gaussian kernels, and generalised by \cite{AL08.1} to reconstruct the
growth rate from the estimated expansion rate. An alternative method
proposed by \cite[][hereafter MB08]{MI08.1} reconstructs the expansion
rate directly from the luminosity-distance data, by expanding them
into a basis system of orthonormal functions, thus avoiding binning in
redshift, and it has been extended in order to estimate the linear
growth factor and to be applied to cosmic shear data (Mignone et al,
in prep.). Also, principal component analysis (PCA) has been used to
reconstruct the dark-energy equation of state parameter as a function
of redshift
\citep[see,~e.g.,][]{HUT03.1,HU05.1,LI05.1,SIM06.1,HU07.2}.

In addition to data interpretation, the last years also saw the
proliferation of several cosmological models based on a very wide
spectrum of physical assumptions, such as for example the existence of
dark energy and dark matter, gravity beyond the standard general
relativity framework or peculiar large scales matter distributions
\citep[for a recent review see][]{DU08.3}. It thus raised the issue of
comparing all these models and to study their mutual degeneracies in
an efficient way which is not straightforward because of their very
different physical backgrounds; this is the case also when different
dark energy prescriptions are adopted.

In this paper, we make use of a principal component approach to define
a basis system capable of providing a parameterisation describing
cosmologies independently of their background physics and of allowing
for the detection of possible unexpected features not foreseen by the
adopted models. For the latter point, this basis system is used to
improve the MB08 method to derive the Hubble expansion rate from
supernova luminosity distances through a direct inversion of the
luminosity distance equation. Our principal component approach differs
from those already proposed in literature because it aims at modelling
observables rather than underlying physical quantities such as
$w$. This is done starting not from data but from theoretical models,
ensuring the derived basis to be optimised for the specific model.

The structure of the paper is as follows. In Sect.~(2) we discuss the
optimal basis system's derivation and properties, in Sect.~(3) we
discuss how the projection on the defined basis system can be used as
a new cosmological parameterisation, in Sect.~(4) we optimise the non
parametric Hubble expansion reconstruction proposed by MB08. Finally
we present our conclusions in Sect.~(5).

\begin{figure*}[!t]
  \centering
  \includegraphics[width=0.48\hsize]{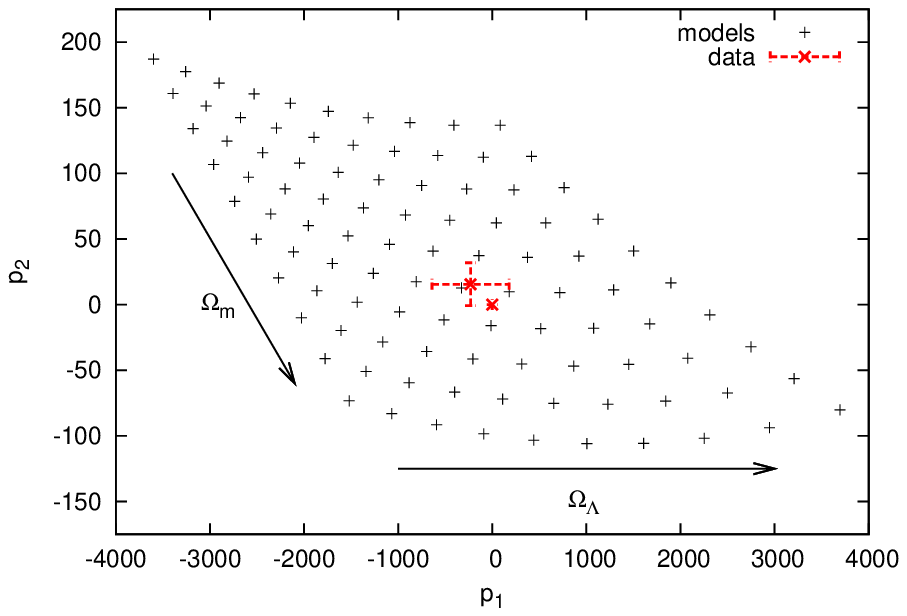}
  \includegraphics[width=0.48\hsize]{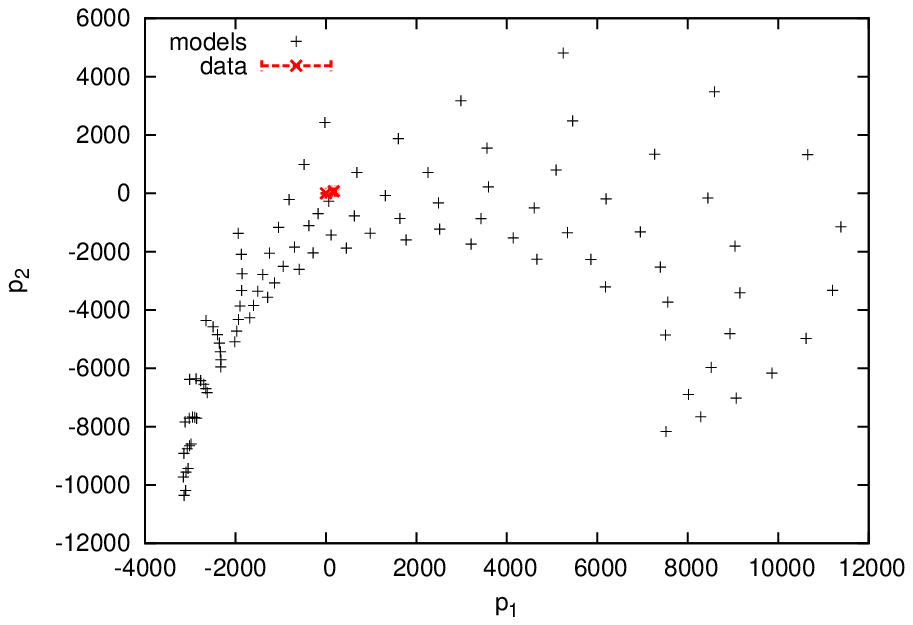}
  \caption{
    The first two features, i.e. the first two components of
    the feature vectors $\vec{\tau}$, of an ensemble of Friedmann
    cosmologies with $h=0.7$, $w_{DE}=-1$, $\sigma_8=0.8$, and the
    matter density and dark energy density ranging in the intervals
    $0.1<\Omega_m<0.5$ and $0.5<\Omega_\Lambda<0.9$, respectively.
    Each black point is related to the corresponding $i$-th
    cosmological model by the mapping $\vec{\tau}_i =
    \vec{W}^T\vec{t}_i$. The left panel refers to SNLS data alone and
    the right panel to WMAP-5yr data alone. The red points mark the
    feature space origin, i.e. the reference model $\bar{\vec{t}}$,
    and the red points with error bars show the projection of the two
    data sets.
  }
  \label{fig:features}
\end{figure*}

\section{An optimised basis system for cosmological data sets}
\label{sec:method}

\begin{table*}[!ht]
  \caption{Summary of the adopted main quantities.}
\label{tb:summary}
\centering
\begin{tabular}{|lll|}
  \hline
  Data set & $\vec{d}\;\in\; \mathbb{R}^{n}$, &~ where ~ $n$ is the number of data points\\
  Training vectors & $\vec{t}_i \;\in\; \mathbb{R}^n$ &~ with ~ $i=1,...,M$\\
  Reference vector/model & $\bar{\vec{t}}\;\in\; \mathbb{R}^{n}$ & ~ e.g. ~ $\bar{\vec{t}}=\langle \vec{t}\rangle$\\
  %%$W:\mathbb{R}^n \rightarrow \mathbb{R}^n$
  Principal components & $\vec{w}_i \;\in\;\mathbb{R}^n$ & ~ with ~ $i=1,...,n$\\
  Feature vector & $\vec{\tau} = \vec{W}^T\vec{t} \;\in\; \mathbb{R}^n$ & ~ where ~ $\vec{W}=\left(\vec{w}_1,\vec{w}_2, ..., \vec{w_n} \right)$\\
\hline
\end{tabular}
\end{table*}

This section presents an application of principal component analysis
to cosmological data sets. In contrast with literature, we do not
search for the principal components describing physical quantities
within a specific cosmological model (e.g. the dark energy equation of
state parameter $w$). Instead, we aim at the principal components
which directly describe cosmological observables (e.g. luminosity
distances, the CMB power spectrum, etc.).
Their derivation does not involve any data set but it is only based on
the predicted behaviour of cosmological models with different
parameters or physical assumptions. The data set may (or not) enter in
a second moment, where it can be analysed by means of the principal
components.

The transformation identified with these principal components is
defined such as to maximise the capability of discerning different
cosmological models and to highlight the possible existence of
unexpected features not foreseen when a specific model is adopted. We
derive our approach having in mind the analysis of cosmological data
sets, but its application is completely general.

\subsection{Principal components derivation}\label{sec:PC_derivation}

We represent any data set with a vector $\vec{d} \;\in\mathbb{R}^n$
whose dimension $n$ corresponds to the number of available data
points, e.g. the number of observed supernovae luminosity distances
$\vec{d}=\left[D_l(z_1), D_l(z_2), ..., D_l(z_n)\right]$ or CMB power
spectrum multipoles $\vec{d}=\left[C(l_1), C(l_2), ...,
  C(l_n)\right]$. This allows one to consider the whole data set as a
single point belonging to an $n-$dimensional space. This
$n$-dimensional space, containing all of these possible $n$-vectors,
makes it possible to address the problem directly through observable quantities
regardless of their underling physics.

To probe the possible observables behaviour, we investigate this
space by populating it with a set of $M$ vectors $ \left\{\vec{t}_i
  \;\in\; \mathbb{R}^n \;\mid\; i=1,...,M\right\}$ modelling the
nature of the observed quantity, e.g. luminosity distances or CMB
power spectrum multipoles, and distributed such as to include the data
set.
Since the ensemble of these models initialises our method, we refer to
it as the {\it training set} in full analogy to other applications of
the same and of similar techniques. In fact, also in morphological or
spectral classification the training set is an ensemble of vectors
sampling the possible behaviours expected in the data. In the same
way, we sample the possible behaviours of cosmological observables to
obtain a basis system capable to distinguish the different underlying
cosmologies. For convenience, we organise the training set in a
matrix,
\begin{equation}
  \vec{T}=\left(\vec{t}_1,\vec{t}_2, ..., \vec{t_M} \right) \;\in\;
  \mathbb{R}^{n\times M} \;,
\end{equation}
where the $\vec{t}_i$ vectors have the same structure of the analysed
data, which can be discrete and irregular as for example the redshift
coverage of a given supernova survey. In principle, these models can
be a set of arbitrary functions, and actually the choice is fully
arbitrary, but it is convenient to consider models at least weakly
resembling the data set. It is in fact pointless to sample the entire
domain of behaviours when we at least know the main data
properties. The choice of the training set only determines for which
kind of models, or better behaviours of the observables, the derived
principal components performance is optimal. This is the reason why we
find convenient to use a set of theoretical models. This choice does
not preclude the method flexibility as it will be demostrated.

Once the training set models are defined, their information content
can be optimised via a linear transformation $W:\mathbb{R}^n
\rightarrow \mathbb{R}^n$ mapping the training-set vectors into a
space (hereafter {\it feature space}) where their projections,
\begin{equation}\label{eq:projection}
  \vec{\tau}_i = \vec{W}^T\vec{t}_i \;\in\; \mathbb{R}^n \;
  \quad\mbox{with}\quad i=1,...M \;,
\end{equation}
have the maximum separation in very few components.
We call {\it feature vectors} any vector resulting from the projection
expressed by Eq.~(\ref{eq:projection}) and {\it features} their
components. The linear transformation, $\vec{W}= \left(
  \vec{w}_1,\vec{w}_2, ..., \vec{w_n} \right)$, satisfying the desired
properties, i.e. concentrating in very few features all information
regarding the differences between the models accounted in the training
set, is given by a set of $n$ orthonormal vectors $\left\{\vec{w}_i
  \;\in\; \mathbb{R}^n \;\mid\; i=1,...,n\right\}$ known as principal
components. The principal components are found by solving the
following eigenvalue problem
\begin{equation}\label{eqn:eigenval}
  \vec{w}_i = \lambda_i\, \vec{S} \vec{w}_i \;,
\end{equation}
and by sorting them in descending order $\lambda_i>\lambda_{i+1}$ to
ensure the largest feature separation in the very first
components. Here
\begin{equation}\label{eqn:scatter_M}
  \vec{S}
  = \vec{\Delta}\vec{\Delta}^T 
  \;\in\; \mathbb{R}^{n\times n} \;
\end{equation}
with $\vec{\Delta}=
\left(\vec{t}_1-\vec{\bar{t}},\vec{t}_2-\vec{\bar{t}}, ...,
  \vec{t}_M-\vec{\bar{t}} \right) \;\in\mathbb{R}^{n\times M}$, is the
so-called {\it scatter matrix}, which encodes the differences (or
scatter) between each training vector $\vec{t}_i$, i.e. a given model,
and the reference vector $\vec{\bar t}$ around which the scatter is
maximised. The reference vector defines the origin of the feature
space and is usually set as the mean of the training set
$\bar{\vec{t}}\equiv\langle \vec{t}\rangle$, but a different
$\vec{\bar t}$ can be used instead, depending on the specific problem
at hand. An interesting choice could be the best fit to a given
cosmological model, so that all other models would be described as its
perturbed states. We summarise in Tab.~(\ref{tb:summary}) all
quantities involved in the principal component derivation.

The principal components derived with our approach provide an optimal
basis system to describe a given cosmological observable for different
cosmologies, as for example luminosity distances to SN Ia if the
training set is constitued by luminosity distance models. Note that
they constitute a full basis system for the training-set cosmologies
only. However, they turn out to be very flexible and even able to
reproduce behaviour not even present in the training-set models as
shown in Sec.~(\ref{sec:independency}).
In this work, we choose to base our training set on Friedmann
$\Lambda$CDM cosmologies with different cosmological parameters, but
of course other kinds of cosmological models can be used as well, such
as e.g.~cosmologies with dynamical dark energy, based on modified
gravity theories, or even a mixture of them so that they can be
optimally described at the same time.

\subsection{Principal components as an optimisation
  problem}\label{sec:PCA_considerations}

The derivation of the principal components can be interpreted as a
constrained optimisation problem, where the subset of linear
orthonormal transformation $\vec{W}$ maximising the separation between
different cosmologies is sought. This is achieved by maximising the
functionals $L_i=\vec{w}_i^TS\vec{w}_i - \lambda_i^{-1}
(\vec{w}^T_i\vec{w}_i-1)$ with respect to $\vec{w}_i$, i.e.~by looking
for the solution of $\delta L_i/\delta {\vec w^T}_i=0$. This leads to
the eigenvalue problem expressed by Eq.~(\ref{eqn:eigenval}) and
consequently to the same principal components $\vec{w}_i$. With this
approach, the first principal component can be seen as an optimal
matched filter which in this case operates directly on cosmological
data sets \citep[see, e.g.][]{SE96.1,MAT04.2,SA04.1}.

\subsection{Speeding up computations}

If the training vector number is smaller than their dimension,
i.e.~$M<n$, only the first $M$ principal components can be associated
to non-vanishing eigenvalues. Therefore, only those components need to
be derived. This is achieved by computing the $M$ eigenvectors
$\vec{w}'\;\in\mathbb{R}^M$ of the matrix
\begin{equation}
  \vec{S}' = \vec{\Delta}^T\,\vec{\Delta}
  \;\in\; \mathbb{R}^{M\times M} \;.
\end{equation}
These are related to the first $M$ eigenvectors of the scatter matrix
$\vec{S}$ by $\vec{w}_i = \vec{\Delta} \vec{w}_i'$. The increase in
computational speed is especially remarkable for large data sets,
where $M\ll n$.

In addition to this gain in computational speed, all the relevant
information is, in most cases, constrained by a very small number of
independent components, $m<M$ (usually up to three for this kind of
applications), allowing for an even stronger dimensionality reduction. In
other words, a full data description is guaranteed by the subspace
$\mathbb{R}^m$ sampled by the training set.

\subsection{Combining different data sets}

The approach described above represents a straightforward way to combine
different observables for a joint data analysis. For example, if we
want to combine the luminosity distances to SN Ia, the CMB angular
power spectrum and the cosmic shear correlation function, we just need
to organise the data vector in the form
\begin{equation}\label{eqn:data_vec}
  \vec{d}=\left[D_l(z_1) \,...\, D_l(z_{n_{sn}}),
    \,C_{l_1} \,...\, C_{l_{n_{cmb}}},
    \,\xi(\theta_1) \,...\, \xi(\theta_{n_\xi})\right] \;,
\end{equation}
whose dimension is given by the sum of all data sets sizes
$n=n_{sn}+n_{cmb}+n_{\xi}$. In order to work with non-dimensional
quantities which reflect the signal-to-noise ratios, the different
observables have to be re-normalised with respect to their variance.
Of course the training set vectors format must be consistent with
Eq.~(\ref{eqn:data_vec}).

\begin{figure*}[!t]
  \centering
  \includegraphics[width=0.49\hsize]{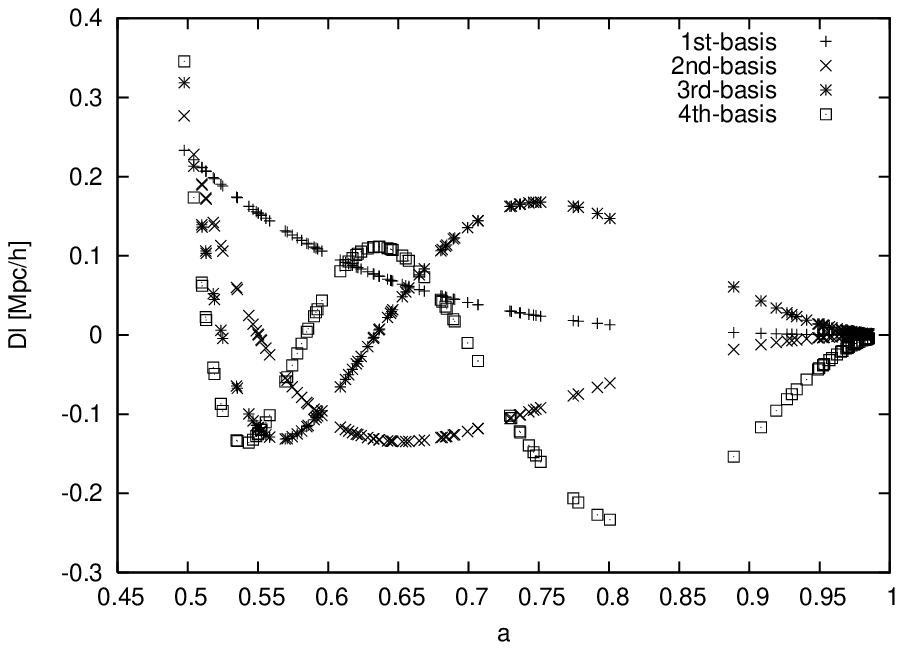}
  \includegraphics[width=0.49\hsize]{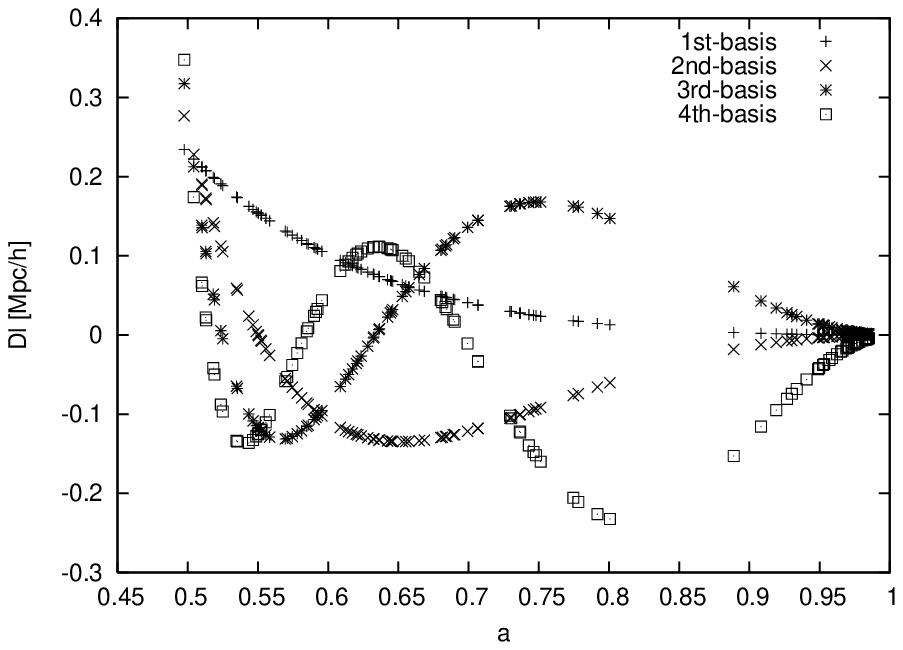}
  \caption{
    Example showing the stability of the principal components
    against the number of models used in the training-set. We show
    only the first 4 principal components derived for the luminosity
    distance sampled at the redshifts covered by the SuperNova Legacy
    Survey. The training-set was produced by sampling the parameter
    space in the range $0.1<\Omega_m<0.5$ and
    $0.5<\Omega_\Lambda<0.9$, 100 (left panel) and 5 times (right
    panel), respectively, and using as a reference cosmology the
    training-set average.
  }
  \label{fig:stability}
\end{figure*}

\section{A new cosmological parameterisation}

Since the features $\vec{\tau}$ discussed in Sec.~(\ref{sec:method})
retain all significant cosmological information, they can be used to
parameterise cosmologies. In contrast with the `standard' cosmological
parameters, they aim to describe observable quantities instead of
physical properties.
To give a visual impression of how cosmologies are represented in the
feature space, we show two simple examples in Fig.~\ref{fig:features},
for luminosity distances only (left panel) and CMB power spectra only
(right panel). Here, we plot the first two components of the feature
vectors resulting from the projection of an ensemble of non-flat
$\Lambda$CDM models. Each point represents a $\Lambda$CDM cosmology
with specific cosmological parameters. The principal components
underlying this projection are based on a training set where only the
matter and dark-energy densities are varied independently in the range
$0.1<\Omega_m<0.5$ and $0.5<\Omega_\Lambda<0.9$, respectively. The
Hubble constant in units of $100{\;\rm km/s/Mpc}$, the equation of state
parameter of dark-energy and the matter fluctuations power spectrum
normalisation were fixed to $h=0.7$, $w=-1$, $\sigma_8=0.8$,
respectively.
Note that, in the case of CMB data, the $\Omega_m-\Omega_\Lambda$
feature space plane is curved such that at least three features would
be necessary for a satisfactory description of the most extreme
cosmologies considered. This is because of the rich complexity of the
data set. To cope with this, a non linear mapping could be used to
`follow the distortion' of the models hyper-plane in the feature
space, but this would add unnecessary complications since the use of a
larger number of features is not a limitation. In any case the actual
CMB physical models have a large number of parameters with large
mutual degeneracies (for instance the optical depth, the baryon
fraction, the inflation spectral index, etc.), therefore the more
complex models usually adopted are not necessarily described by a
linearly growing number of features compensating the increase of
necessary features.

With this formalism, the principal components can be considered as
cosmological eigen-modes ({\it eigen-cosmologies}) where observations
would ``excite'' (i.e. make visible) a given number of modes according
to their accuracy.

\subsection {The advantages}\label{sec:advantages}

The use of these orthonormal functions to define a parameter set
characterising observable behaviour instead of underlying physical
quantities has several advantages. In fact:
\begin{itemize}
\item the features are fully independent by definition and therefore
  avoid any redundancy and degeneracy in the observable description,
  in contrast with physical parameterisations;
\item they retain all available information because they are derived
  from the principal components;
\item their number is minimal as allowed by the data accuracy;
\item they provide the best discriminatory power for the family of
  cosmologies adopted in the training-set;
\item they can be related to any physical model via the mapping
  $\vec{\tau}_i = \vec{W}^T\vec{t}_i$ itself;
\item the features allow one to quantify the overall difference between
  two cosmologies in terms of a simple metric separation
  \begin{equation}
    s=\frac{\left|\vec{\tau}_1-\vec{\tau}_2\right|}
           {\left|\vec{\tau}_\sigma\right|} \;,
    \label{eq:separation}
  \end{equation}
  where $\vec{\tau}_1$ and $\vec{\tau}_2$ are the two cosmologies
  features vectors and $\vec{\tau}_\sigma$ is the data uncertainty
  projection in the feature space.
\end{itemize}

These properties apply also to cosmologies which are not explicitly
included in the training set; however, in this case not all model
behaviours are ensured to be captured. In other words, if nature or
the cosmological model we are investigating differs from the one
adopted in the principal component definition, we could still use
them, even if in suboptimal conditions. In any case, it is possible to
cope with this by making the training set less specialised. If we are
for example studying cosmologies based on different physical
frameworks such as General Relativity, TeVeS or $f(R)$ theories or
simply different dark-energy models, we could include all of them in
the training set so that the resulting features can optimally describe
all of them at the same time. Given that, it follows how the features
$\vec{\tau}$ can be used as a common parameterisation to describe and
compare cosmologies even if based on different physical
frameworks. Again, this is possible because this approach parameterises
observables only and not their very diverse background physics. In
this paper we only consider non-flat $\Lambda$CDM cosmologies for sake
of simplicity.

\subsection{Studying different modellisations degeneracies}

The proposed parameterisation, thanks to the properties discussed in
Sec.~(\ref{sec:advantages}), provides a useful tool to study
degeneracies in the same or, more interestingly, in different
modellisations and physical frameworks. In fact, fully degenerate
models show the same observational properties and consequently have
the same features $\vec{\tau}$. Of course, when considering
observational data, degeneracies are not associated to a feature space
point but to the hyper-volume defined by the data errors projection.
In fact also data errors have to be projected into the feature space
to define the region compatible with the data as shown in
Fig.~(\ref{fig:features}). All information regarding how similar,
i.e. degenerate, two models are is quantified by the metric distance
given in Eq.~(\ref{eq:separation}) which is in fact normalised with
respect to the data accuracy. In fact, all models whose separation is
smaller than the hyper-volume radius allowed by data are degenerate.

In comparison with Markov-Chain Monte Carlo methods \citep[see for
example][]{LE02.1}, this approach is not an iterative method and is
computationally cheap since a small number of models have to be
computed. In fact, the parameter space can be sampled on a very coarse
grid and, if necessary, according to the parameters conditional
distribution in analogy with Gibbs sampling. In a follow up paper we
will discuss a detailed study of this parameterisation and of its
application in degeneracy studies.

\section{Hubble expansion rate from supernovae data}
\label{sec:hubblerate}

Supernova luminosity distances are a very powerful probe to
investigate cosmology. In particular, they can be used to directly
measure the expansion history of the universe, $H(a)$, avoiding any
reference to Friedmann models. In fact, if we assume a topologically
simply connected, homogeneous and isotropic universe, the luminosity
distance can be expressed as
\begin{equation}\label{eqn:lum_distance}
  D_L(a)=\frac{c}{H_0\;a}\int_a^1\frac{\d x}{x^2}e(x) \;,
\end{equation}
where the expansion function is expressed as $H(a)=H_0E(a)$, with
$H_0$ being the Hubble expansion constant and $e(a)\equiv E^{-1}(a)$
the inverse expansion rate. For the sake of simplicity, we drop the
$c/H_0$ factor in the following discussion.

The Hubble expansion function can be directly derived from a
luminosity distance data set, as detailed in MB08. In fact, the
derivative of Eq.~(\ref{eqn:lum_distance}) with respect to the scale
factor, $a$, can be brought into the form of a Volterra integral
equation of the second kind,
\begin{equation}
  e(a)=-a^3D'_L(a)+a\int_1^a\frac{\d x}{x^2}e(x) \;,
\end{equation}
whose solution, $e(a)$, can be expressed in terms of a Neumann series
\citep[see for e.g.][]{AR95.1},
\begin{equation}
  e(a)=\sum_{i=0}^\infty a^if_i(a) \;,
\end{equation}
where a possible choice for the expansion terms $f_i$ is
\begin{equation}
  f_0(a)=-a^3D'_L(a) \;,\quad f_i(a)=\int_1^a \frac{\d x}{x^2} f_{i-1} \;.
  \label{eq:ebase}
\end{equation}
Here it is necessary to smooth the observational data, $D_L(a)$, to
avoid possible issues with the intrinsic data scatter when estimating
the derivative $D'_L(a)$. A convenient way to do it is to expand the
luminosity distance data into a set of orthonormal functions
\begin{equation}\label{eq:expansion}
  D_L(a)=\sum_{j=0}^Mc_jp_j(a) \;,
\end{equation}
where the coefficients $c_j$ are determined from data fitting. This
approach has the advantage of avoiding all assumptions regarding the
energy content of the universe, thus making the reconstruction nearly
model-independent.

The choice of the adopted basis, $\{p_j\}$ in
Eq.~(\ref{eq:expansion}), is arbitrary. For illustrative reasons, MB08
adopted the linearly independent set $u_j(a)=a^{j/2-1}$
ortho-normalised with the Gram-Schmidt process.  However, the basis
$\{p_j\}$ can be defined such as to minimise the number of necessary
modes and to have them ordered according to their information
content. A good choice fulfilling these criteria is represented by the
principal components defined in Sec.~(\ref{sec:method}) which can be
optimised for a specific cosmology or for a set of cosmological models
based on different physical assumptions. This basis optimisation
enhances the MB08 method performances without precluding its
flexibility. In fact, also behaviours not described by the models
adopted in the basis definition can be reproduced, as it will be shown
in Sec.~(\ref{sec:independency}).

\begin{figure}[!t]
\centering
\includegraphics[angle=270,width=0.95\hsize]{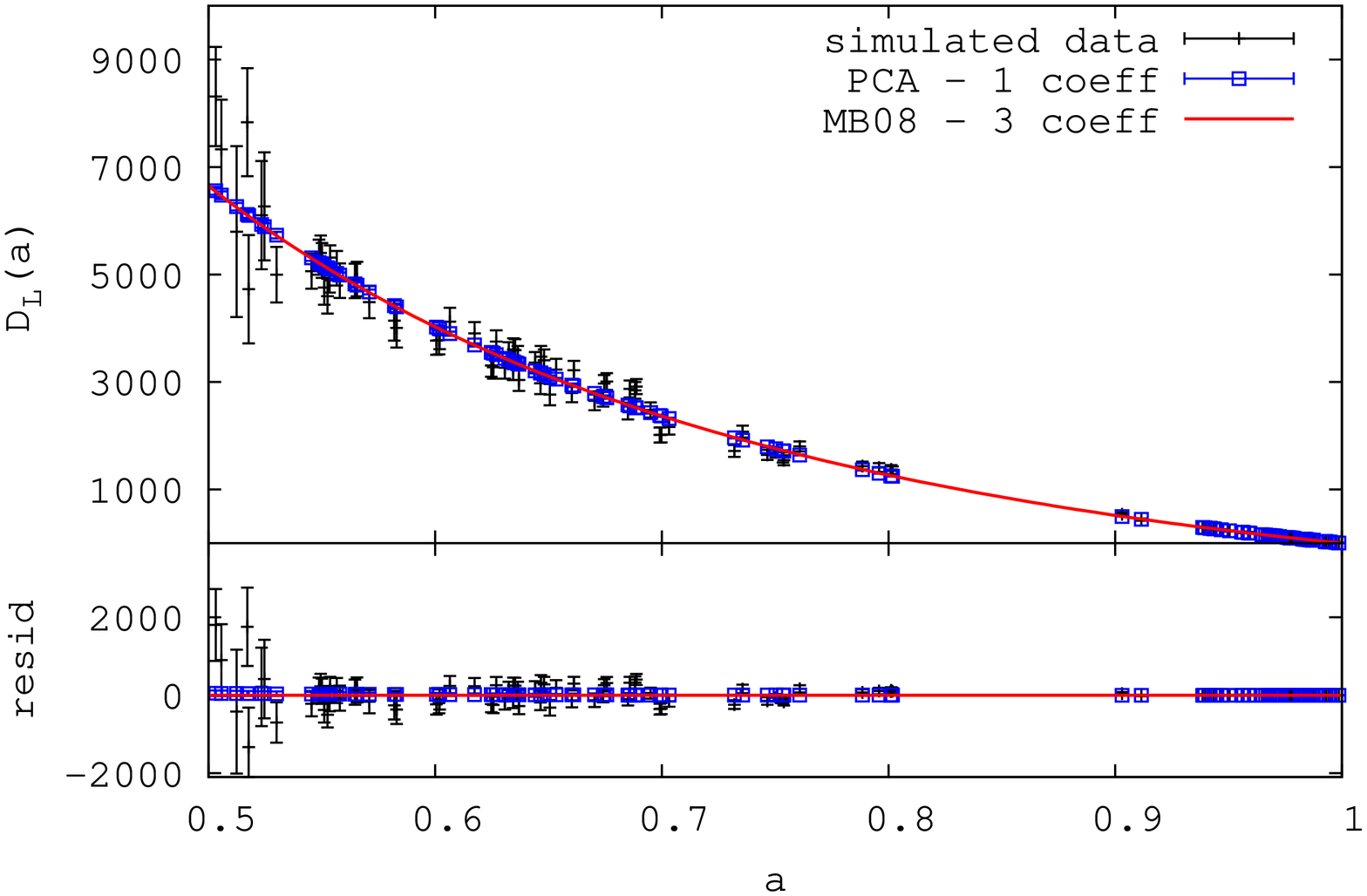}
\includegraphics[angle=270,width=0.95\hsize]{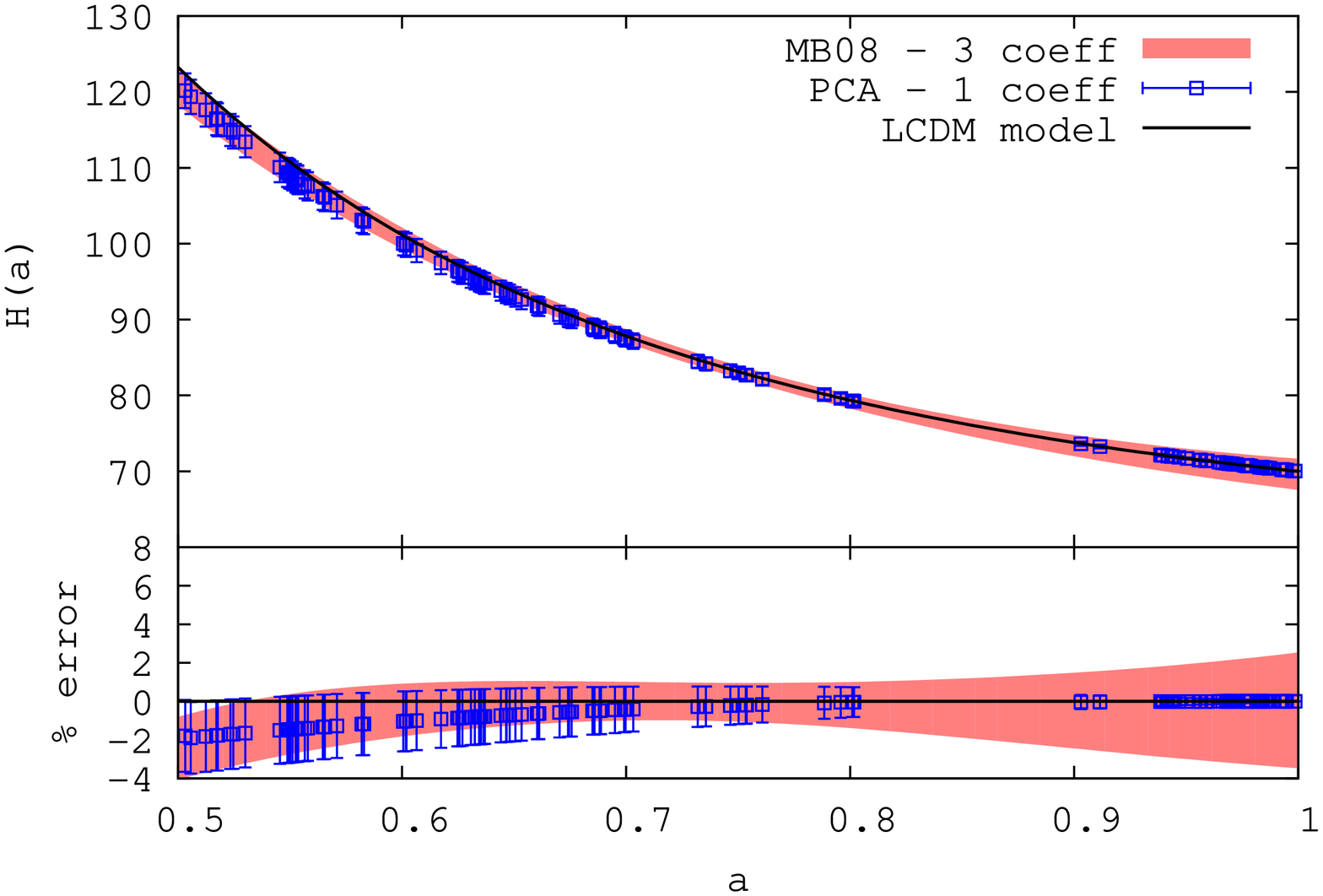}
\caption{
  Fit of our supernovae luminosity distances mock catalogue
  resembling the SNLS data set (left panel) and the recovered
  expansion rate obtained (right panel). We compare the results
  obtained by using the original MB08 recipe (shaded area) and by
  using our principal components as basis set (blue squares).  The
  increased accuracy is especially evident at lower redshifts, where
  the method improvement takes fully advantage of the smaller
  measurement errors.
}
\label{fig:mb08_cf_all}
\end{figure}

\subsection{Principal components stability}\label{sec:stability}

The stability of the principal components with respect to the number
of models used in the training set has been tested.  We show in
Fig.~\ref{fig:stability} the first four principal components derived
for a luminosity-distance data vector with the same redshift sampling
of the SuperNova Legacy Survey \citep{AS06.1}.
The training set is based on non-flat $\Lambda$CDM models with
$h=0.7$, $w=-1$ and the matter and dark-energy density parameters
sampling the ranges $0.1<\Omega_m<0.5$ and $0.5<\Omega_\Lambda<0.9$,
respectively; as a reference cosmology, the average of the
training set has been used. The ($\Omega_m,\Omega_\Lambda$) space was
regularly sampled by the training set $10,000$ times to produce the
left panel and only $25$ times in the right panel. Clearly, the
principal components are very stable against the training-set size and
only depend on the range spanned by the cosmological parameters of the
training set.

As discussed in Sec.~(\ref{sec:method}), the information content of
each principal component is quantified by the corresponding
eigenvalue, which in this case are $\lambda_1=1$,
$\lambda_2=2.0\,10^{-4}$, $\lambda_3=1.4\,10^{-7}$ and
$\lambda_4=1.2\,10^{-10}$. Hence, all information and discriminatory
power is concentrated in the very first components allowing for a strong
dimensionality reduction, from $n=117$ (i.e. the number of supernovae
in the data set) to 1 or 2 dimensions for this specific case. If the
number of parameters sampled in the training-set construction is
increased, or if the intervals over which they are sampled are larger,
the power is distributed towards higher orders, but is still fairly
concentrated in very few components.

\subsection{Application to synthetic data: highlighting unexpected
  features}\label{sec:independency}

We apply the MB08 method combined with the principal components
described in Sec.~(\ref{sec:method}) on a synthetic data sample drawn
from a $\Lambda$CDM model with $\Omega_M=0.3$, $\Omega_{\Lambda}=0.7$
and $h=0.7$, and resembling the SNLS properties \citep{AS06.1}. The
training set for the principal components definition is the same
tested in Sec.~(\ref{sec:stability}).
We show in Fig.~\ref{fig:mb08_cf_all} the resulting fit to the data
(left panel) and the subsequently estimated expansion rate (right
panel) as compared with the original MB08 recipe (shaded area). The
resulting error bars are smaller when the principal components are
used since the adopted basis is optimised for $\Lambda$CDM cosmologies
and less parameters have to be fitted: one coefficient, rather than
three. The increased accuracy is particularly evident at lower
redshifts, where the method improvement takes fully advantage of the
smaller measurement errors.

\begin{figure}[!t]
\centering
\includegraphics[angle=270,width=0.95\hsize]{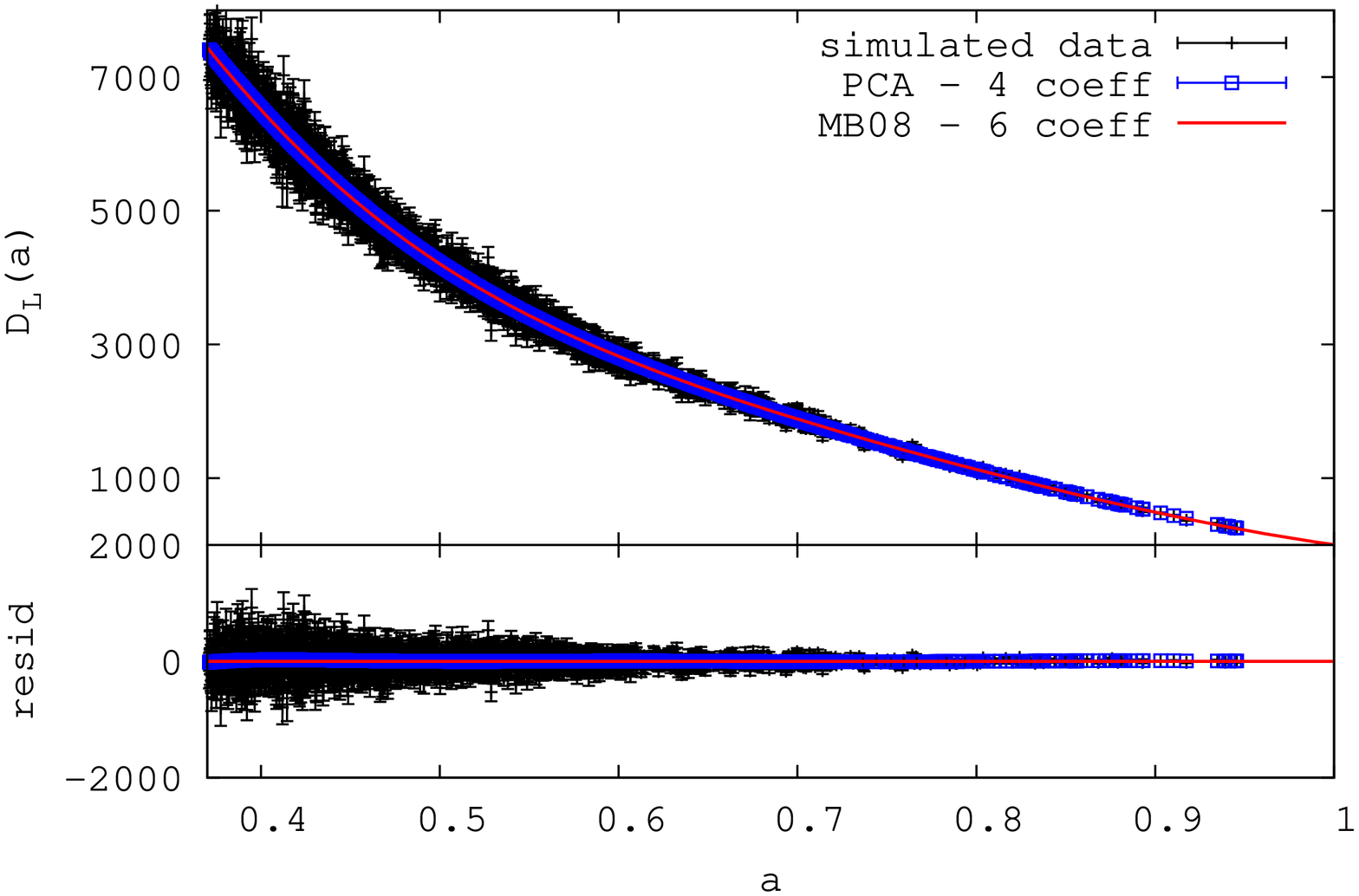}
\includegraphics[angle=270,width=0.95\hsize]{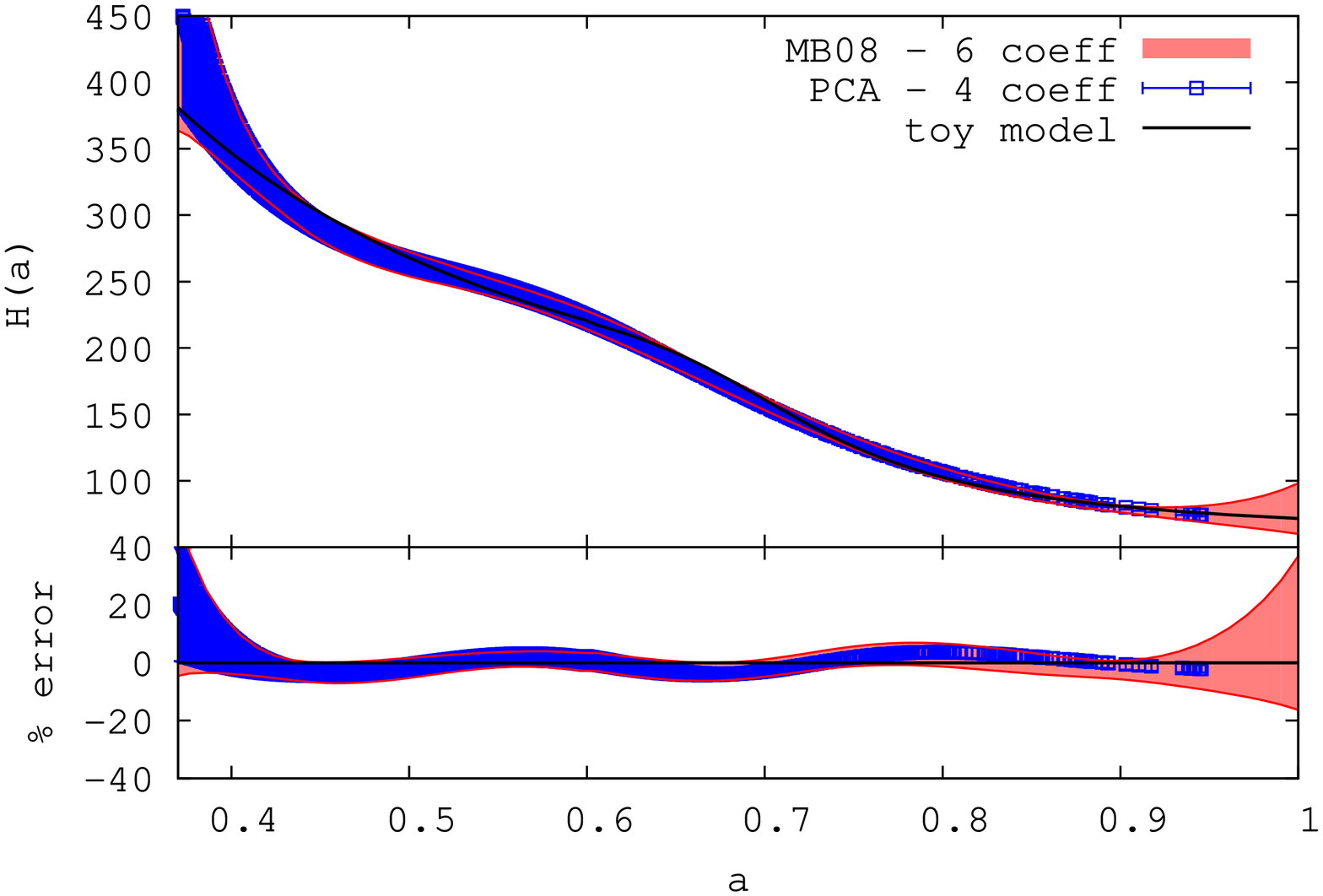}
\caption{Same as in Fig.~\ref{fig:mb08_cf_all}, but for a SNAP like
  mock catalogue based on a toy-model simulation with a sharp
  transition in the expansion rate. Even if the training-set was
  defined on Friedmann models without such a feature, the
  reconstruction was capable to highlight it.}
\label{fig:toymodel}
\end{figure}

As a second test case, we now consider a more challenging data set in
order to test the method's capability of capturing behaviours not
explicitly described by the training-set models. In this example, we
use the same training set of the previous case but we analyse
luminosity distances resulting from a toy-model cosmology with a
sudden transition in the expansion rate (see MB08 for details). In
this case, the sample observational characteristics are modelled after
the proposed satellite SNAP \citep{AL05.1}.
A $\chi^2$-analysis shows that this simulated data are compatible with
a standard Friedmann $\Lambda$CDM cosmology. This is of course a
misleading result since the background cosmology has a completely
different nature and the sudden transition in $H(a)$ is not
highlighted. This demonstrates how a standard $\chi^2$-approach is not
always capable of revealing the unexpected features hidden in the data
set and how the result is bound to the theoretical prejudice.  In
contrast, with the proposed method this expansion rate transition is
observed even if the training set does not contain any model with such
a feature. The reconstructed expansion rate for this example is shown
in Fig.~\ref{fig:toymodel} (blue error bars) together with the one
obtained with the method as originally proposed by MB08 (shaded area).
The accuracy improvement with respect to the original MB08 method may
not appear striking, but we have to consider that this is a very
extreme case where not even the best-fit Friedmann model is covered by
the training set. This example just demonstrates the method's
capability of capturing unexpected features even if optimised only for
a specific set of cosmologies.

\subsection{Results on the SNLS data set}

\begin{figure}[!t]
  \centering
  \includegraphics[angle=270,width=0.95\hsize]{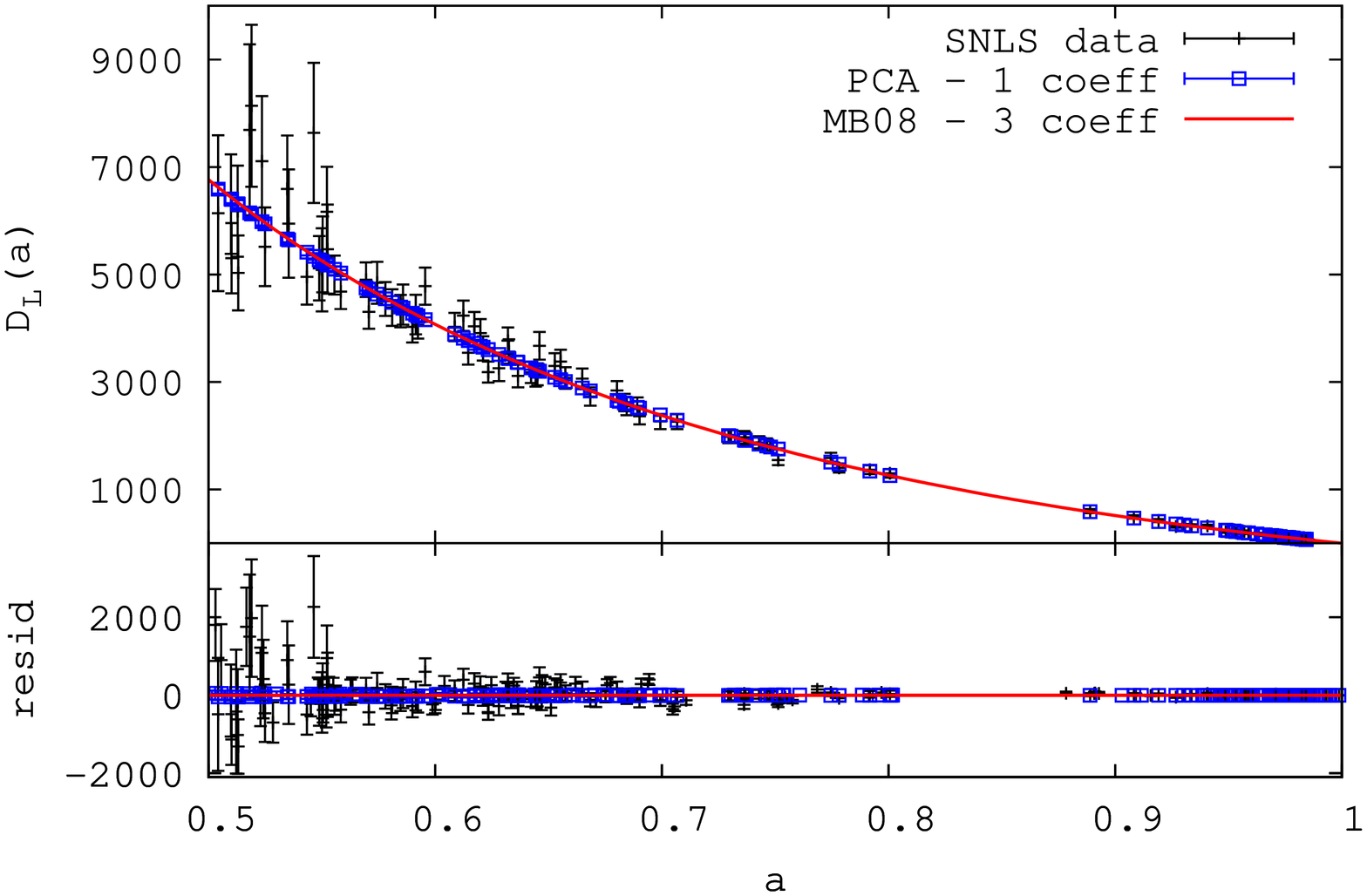}
  \includegraphics[angle=270,width=0.95\hsize]{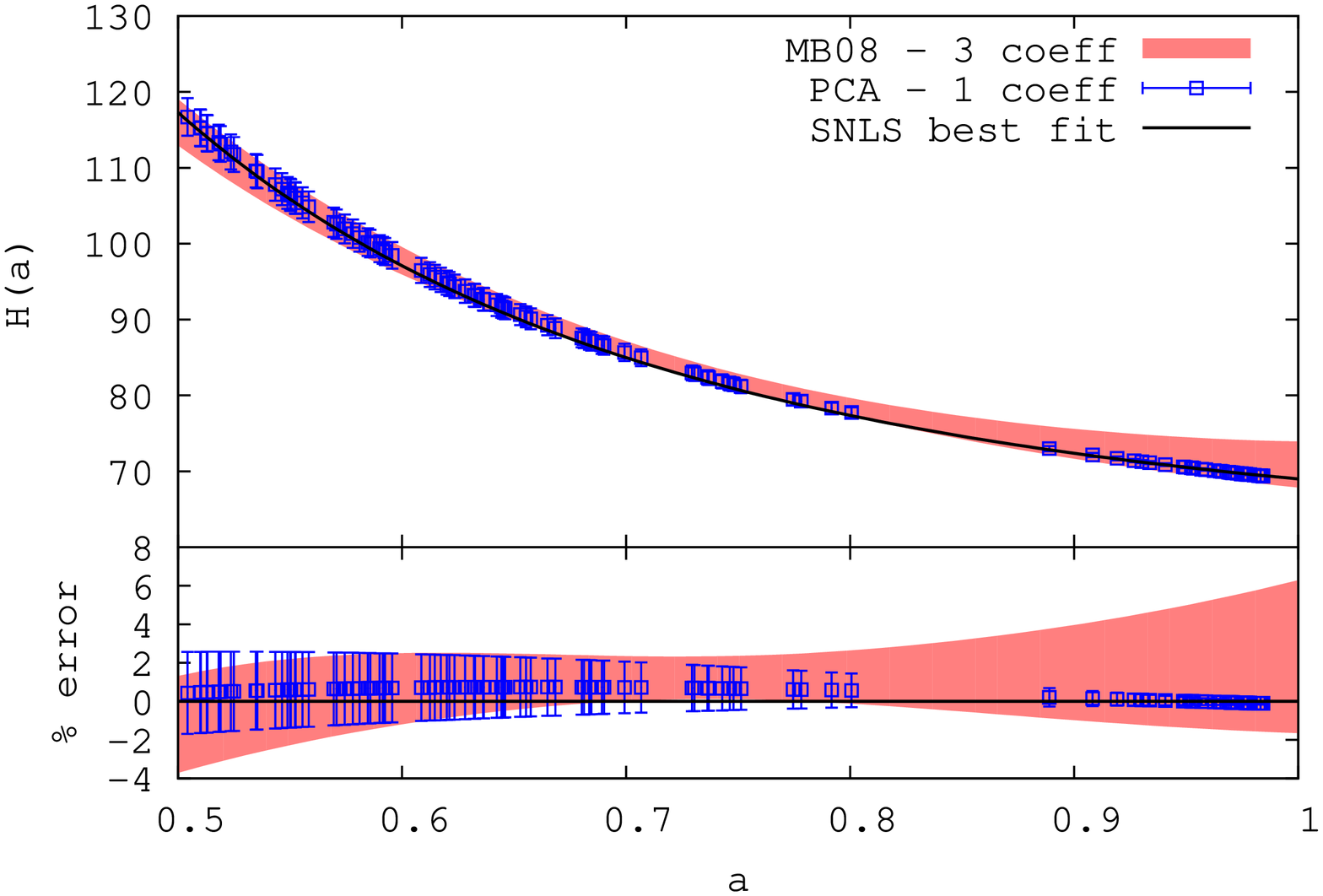}
  \caption{Same as in Fig.~\ref{fig:mb08_cf_all} but for the real SNLS
    data set. The reconstruction accuracy is largely improved and is
    fully compatible with the best-fit $\Lambda$CDM model.}
  \label{fig:snls}
\end{figure}

The SuperNova Legacy Survey data set consists of 118 supernovae in the
redshift range $0.015<z<1.01$, 71 observed with the
Canada-France-Hawaii Telescope and 44 taken from the literature
\citep{AS06.1}. We analysed this sample with the same procedure
applied to the $\Lambda$CDM simulation discussed in
Sec.~(\ref{sec:independency}). The result shown in Fig.~\ref{fig:snls}
is fully compatible with the best $\Lambda$CDM model fit
\citep{AS06.1}. The accuracy is largely improved with respect to the
original MB08 reconstruction, giving the hope that future supernova
samples may be able to reveal the dark-energy nature.

\section{Conclusions}

We defined an optimal basis system based on principal components to
decompose cosmological observables. The principal components are
defined starting not from the data but from an ensemble of given
models. The basis system can be optimised for a specific cosmological
model or for an ensemble of models even if based on different physical
hypotheses. We suggest two main applications: (1) to define a
cosmological parameterisation applicable to any model independently of
the physical background and (2) to optimise the MB08 method for a
direct estimate of the expansion rate from luminosity distance data.

On one hand, the cosmological parameterisation is based on the
coefficients, i.e. the features, resulting from the observables
projection on the discussed basis system. The features are fully
independent, avoiding the degeneracies and redundancies of physical
parameters, and their number is minimal with respect to the data
accuracy. Since they quantify observable properties, they can be used
as a common parameterisation to describe cosmologies independently of
their background physics. However, they can be uniquely related to
physical parameters once a model is specified. In addition, this
parameterisation allows one to quantify the differences between different
cosmologies in terms of a simple metric separation.

On the other hand, the method proposed by MB08, which directly
estimates the expansion rate from supernova data, requires to expand
luminosity distances into a set of arbitrary orthonormal functions.
The use of the basis system derived in this work largely
reduces the resulting uncertainties and, even if the method is only
optimised for a single or for an ensemble of cosmological models, it
is still capable to detect unforeseen features not included in the
algorithm setup as demonstrated.

\acknowledgements{ It is a pleasure to thank M. Bartelmann for
  carefully reading the manuscript and providing useful comments. We
  also wish to thank L. Verde for helpful discussion. MM is supported
  by the Transregional Collaborative Research Centre TRR 33 of the
  German Research Foundation, CM is
  supported by the International Max Planck Research School (IMPRS)
  for Astronomy and Cosmic Physics at the University of Heidelberg. }

\bibliographystyle{aa}

\end{document}